\documentclass[aps,twocolumn,superscriptaddress,showpacs,floatfix]{revtex4}
\pdfoutput=1
\usepackage{graphicx}

\begin{document}

%\preprint{}
\title{FFTS readout for large arrays of Microwave Kinetic Inductance Detectors}
\date{\today}

\author{S. J. C. Yates}
\email[e-mail: ]{s.yates@sron.nl}
\affiliation{SRON, Sorbonnelaan 2, 3584 CA Utrecht, The Netherlands}
\author{A. M. Baryshev}
\affiliation{SRON, Sorbonnelaan 2, 3584 CA Utrecht, The Netherlands}
\affiliation{Kapteyn Astronomical Institute, University of Groningen, P.O. Box 800, 9700 AV Groningen, The Netherlands}
\author{J. J. A. Baselmans}
\affiliation{SRON, Sorbonnelaan 2, 3584 CA Utrecht, The Netherlands}
\author{B. Klein}
\affiliation{Max-Planck-Institut f\"{u}r Radioastronomie, Auf dem H\"{u}gel 69, 53121 Bonn, Germany}
\author{R. G\"{u}sten}
\affiliation{Max-Planck-Institut f\"{u}r Radioastronomie, Auf dem H\"{u}gel 69, 53121 Bonn, Germany}

\begin{abstract}
Microwave Kinetic Inductance Detectors (MKIDs) have great potential for large very sensitive detector arrays for use in, for example, sub-mm imaging. Being intrinsically readout in the frequency domain, they are particularly suited for frequency domain multiplexing allowing $\sim$1000s of devices to be readout with one pair of coaxial cables. However, this moves the complexity of the detector from the cryogenics to the warm electronics.
We present here the concept and experimental demonstration of the use of Fast Fourier Transform Spectrometer (FFTS) readout, showing no deterioration of the noise performance compared to low noise analog mixing while allowing high multiplexing ratios.

\end{abstract}

\maketitle 
%\emph{INTRO}

The Far-Infrared (FIR) and sub-mm wavelength bands (10\,--\,1000\,$\mu$m) contain a plethora of information on the evolution and formation of galaxies, stars and planetary systems. Future FIR and sub-mm space missions, such as SPICA~\cite{SPICA} and Millimetron~\cite{Millimetron}  require large arrays of detectors with an unprecedented high sensitivity, expressed in a Noise Equivalent Power (NEP)\,$<2\times 10^{-19}$W/Hz$^{1/2}$. Also, many ground based telescopes like IRAM~\cite{IRAM}, APEX~\cite{APEX} and future observatories like CCAT~\cite{CCAT} require very large arrays for faster observation speeds and larger instantaneous fields-of-view.

Microwave Kinetic Inductance Detectors MKIDs~\cite{Day03} are a rapidly developing new detector technology~\cite{Gao07, BarendsPRL08, BarendsPRB09} that has the potential to be used to fabricate very large imaging arrays for the FIR, sub-mm~\cite{Caltech}, optical and x-ray~\cite{MazinAPL06}. The main advantages of MKIDs over other detector technologies are: i) ease of fabrication, ii) the intrinsic adaption for frequency domain multiplexing (FDMUX) using microwave readout signal frequencies and iii) operation with a wide dynamic range. Currently MKIDs have shown sufficient sensitivities in a lab environment for both ground-based and space-based instruments~\cite{Jochem08} and there has been an initial demonstration at the CSO telescope on Mauna Kea, Hawaii~\cite{Caltech}.

However, FDMUX of MKIDs has yet to be satisfactorily demonstrated, particularly anywhere near the (power and weight) requirements needed for space. Previous measurements have been non-scalable using either analog mixing to readout one MKID, which has proven useful in single detector characterization in a laboratory environment, or digital mixing using specialist chips~\cite{Mazin06}.  
Using the latter technique a readout of 16~MKIDs has been demonstrated using one commercial demultiplexer card. However, the technique is hard to scale to very large pixel numbers~\cite{Mazin06}.
In this Letter we present a new MKID readout scheme based upon frequency division multiplexing and a digital Fast Fourier Transform Spectrometer (FFTS)~\cite{Klein06, Klein08}. 
Digital FFTS systems are presently being used as back-end electronics for heterodyne mixers on several ground based telescopes~\cite{Klein08}. They can process a real time data stream with a bandwidth of up to 1.8~GHz into a frequency spectrum of $\sim$~8000 points. We propose here to use the FFTS to read out a set of single frequency probe signals that have passed through a chip containing one MKID detector per probe signal. The proposed readout is able to read out $\sim$~1000 pixels using 1 probe signal generator and 1 FFTS board. To demonstrate the principle, we describe an experiment where we have read out 8 MKIDs simultaneously and we have measured both the dark detector NEP and the response to an optical signal for 8 MKIDS. These results are compared to a conventional analog single MKID readout scheme and show that the FFTS based multiplexed  readout does not deteriorate the intrinsic system noise of our setup, which is limited by the first stage cryogenic amplifier.

%\emph{KID general}

MKIDs are superconducting pair breaking detectors that sense the change in the complex surface impedance of a thin superconducting film due to radiation absorption with a (sky) frequency F$_{rad}>2\Delta/h$, which is $\sim$~80~GHz for aluminum. An MKID consists of a thin superconducting film that is incorporated in a resonance circuit which is either capacitively~\cite{Day03} or inductively~\cite{Doyle08} coupled to a through line. Changes in the surface impedance of the film are converted to changes in resonator quality factor and resonance frequency. These changes can be read out by measuring the phase and amplitude modulation of a probe tone at a probe frequency equal or close to the resonator resonance frequency $F_0$, which is typically a few GHz.

MKIDs take advantage of the fact that a superconductor at Tc/10 has negligible losses at the probe frequency $f_0$. This enables very low 3~dB bandwidth $\delta f$ resonators with high Q factors $f_0/\delta f\sim 10^6$. Hence, close packing of the resonators in (readout) frequency space is only limited by manufacturing tolerances ($\sim$1~MHz at $f_0\sim$\,3\,GHz). For example, using conventional one octave bandwidth amplifier (4-8~GHz) one can read out $\sim$~4000~MKIDs using just one pair of coaxial cables.

\begin{figure}
\includegraphics*{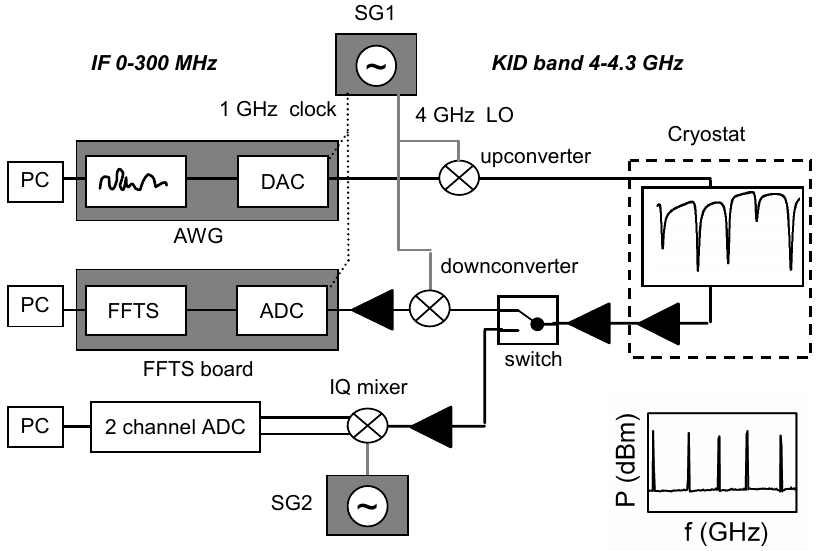}
\caption{The circuit diagram for test in this experiment. The readout signal is generated in software from the known MKID resonance frequencies in the IF (0.1\,--\,300\,MHz) and converted to analog by the DAC of the AWG card; the signal is then mixed with signal generator 1 (SG1) up to $\sim$4\,--\,4.3\,GHz using an upconverter (inset lower right), passes through the MKIDs, amplified and down converted back to the IF band.
A last amplification step follows before digitization and processing by the FFTS. For comparison, by using a microwave switch we can switch from the FFTS readout to a conventional single pixel readout scheme using an IQ mixer and a separate LO (SG2) at the resonance frequency of one of the MKIDs.}\label{fig:circuit}
\end{figure}

%\emph{SETUP description}

To demonstrate the FFTS based readout of MKIDs we use the readout scheme  shown in Fig.~\ref{fig:circuit} to readout 8 MKIDs. A set of 8 Intermediate Frequency (IF) probe tones in a band from 0.1\,--\,300\,MHz are generated by a 1\,GSPS ($1\times10^9$ samples per second) commercial arbitrary waveform generator (AWG) with a 12-bit Digital-to-Analog Converter (DAC). 
Subsequentially they are mixed to the required readout at Radio Frequency (RF) of 4\,--\,4.3\,GHz by a commercial single sideband upconverter using an Local Oscillator (LO) frequency of 4~GHz from SG1 (example tones shown in the inset Fig.~\ref{fig:circuit}). 
The resulting RF probe signal consists of a set of RF frequency probe tones, where each tone corresponds to one MKID while additional blind tones can be added to measure the system noise.
The LO of the mixer is a low noise synthesizer which also has an 1\,GHz clock output used to clock the DAC and FFTS. The tones are run at as high a level as possible to use the full scale of the DAC and minimize the level of the LO leakage into the RF signal. 
Subsequently, the signals are attenuated using microwave attenuators at room temperature, 4~K and 1~K to the required level MKID readout power of $\sim-90$~dBm per MKID. 
The signals pass through the MKID chip that is cooled to a temperature of 100~mK. The tones are modulated by the MKIDs, amplified at 4~K by a low noise amplifier ($T_n=4$~K) and passed out of the cryostat. 
Here a microwave switch enables the use of the FFTS readout or a conventional single tone readout (IQ readout)~\cite{Jochem08} of one of the 8 multiplexed MKIDs. 
For the single tone IQ readout, the RF probe signal is passed to an IQ (quadrature) mixer where the LO is generated by a second signal generator (SG2) tuned to the exact probe tone frequency of one of the resonators. 
The two signal generators, DAC card and FFTS board are phase locked using a 10 MHz reference clock to reduce the relative clock jitter. By changing the SG2 frequency we can switch between all 8 probe tones. The IQ mixer outputs the In-phase (I) and Quadrature-phase (Q) signals of the tone of the selected MKID at baseband that is then digitized using a dual channel 16 bit ADC at 2$\times10^5$ samples/s.

For the multiplexed FFTS readout, the RF probe signals go to a downconverter which is driven by the same LO as the upconverter (SG1). As a result the RF probe tones are converted back to the exact same IF frequencies as the original IF tones. A final amplification is used to compensate for conversion loss at the mixer to efficiently use the full scale deflection of the FFTS Analog to Digital Converter (ADC), which digitizes the signal at 8 bit and 2~GSPS.
The FFTS performs a continuous N=8192 bin (point) complex Fast Fourier Transform (FFT) outputting a N bin power spectrum for every block of N points of time domain data. 
The outputted 8192 bin spectrum has a binwidth of 1~GHz/8192=122~kHz, where the value of each bin is the summed spectral power in the bin from that block. 
By comparing the spectrum from different blocks of time domain data we can follow the time evolution of the spectral power in each bin. 
By arranging that the probe frequencies occupy different bins this one board will be equivalent of 8192 digital downconverters~\cite{note:bw}.
A selection of the bins have a probe tone in them allowing the time evolution of the probe tones to be simultaneously followed at the FFT rate of 122~kHz, intrinsically multiplexing the readout of the tones in the frequency domain. 
Note that we measure now the modulation of the probe tone power which is a measure of the power transmission of the MKIDs, $|S_{21}|^2$.
The signal bandwidth is then reduced by averaging multiple spectra for 20~ms or more. 
This final integration bandwidth sets the noise bandwidth of the readout, not the bin bandwidth.

The devices used in the experimented  are coplanar waveguide quarter wavelength resonators~\cite{Day03,Jochem08} capacitively coupled to the through line. The devices are made from a single 100\,nm thick Al film thermally evaporated on a high purity silicon substrate. The measured chip has 25 usable MKIDs  at slightly different resonance frequencies. 8~MKIDs with Q $\sim$\,150000 were chosen for the experiment. 
\begin{figure}
\includegraphics*{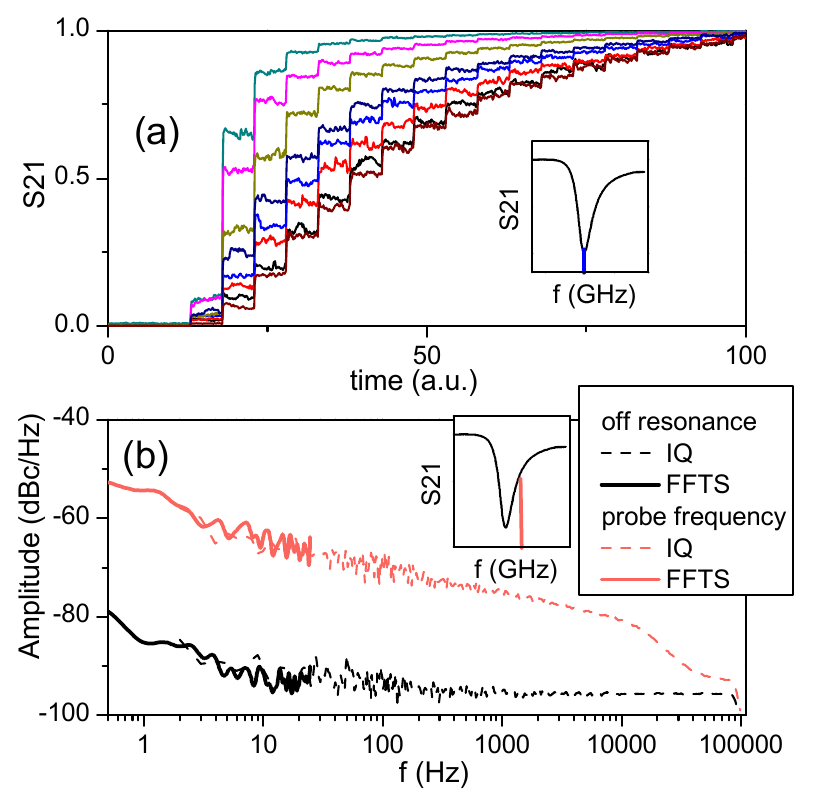}
\caption{(a): Measured response of 8\,MKIDs in time to a optical step ramp from a red LED. The initial probe frequency was on resonance, the LED was stepped linearly in current. (a inset): $|S_{21}|^2$ transmission curve versus measurement frequency measured by the FFTS technique of a MKID around the resonance frequency. The frequency was swept by changing the LO frequency of the mixers, keeping the DAC output constant. Indicated are the approximate probe point and the off resonance frequencies. (b): Power spectral density of noise compared to carrier in the signal amplitude at off resonance (lower) and the probe point which was 8\,kHz from resonance (upper), measured using the FFTS readout and compared to measurements using an IQ mixer.}\label{fig:compoundfig}
\end{figure}

%\emph{exp results}

In the first experiment we measure the response of 8 probed MKIDs to a red LED illumination of the chip with optical photons, increasing the LED current in discrete steps~\cite{note:LED}. The insets of Fig.\,\ref{fig:compoundfig} show the measured frequency dependent transmission and probe frequencies for one MKID. 
Fig.~\ref{fig:compoundfig}(a) shows the multiplexed (i.e. FFTS readout) response of 8~MKIDs with the probe tones set on resonance when there was no LED illumination. We clearly observe the expected step wise increase in $|S_{21}|^2$. Non uniform illumination of the chip is the reason of observed response difference. 

%\emph{Exp results II}

To fully demonstrate the readout we need to study not only the MKID response but also we need to study the noise properties of the new readout technology. 
To do this, we sample $|S_{21}|^2$ of each probe signal using the FFTS at a rate of one measurement every 20~ms for 10~s. 
In this experiment we have used probe tone frequencies at $\sim 0.6 \delta f$ away from the resonance frequency. 
The measurement is repeated by tuning the probe signal 1~MHz off resonance, effectively measuring the system noise not influenced by the MKID. 
This data is used to calculate the power spectral density with respect to the carrier amplitude, the result is shown for one representative of the 8 MKIDs by the bold lines in the figure, both for 8~kHz and 1~MHz off resonance. 
Subsequently, we measured the noise spectrum for each MKID separately in time both at 8~kHz and 1~MHz off resonance by using our conventional IQ setup. The result is shown by the thin lines in the Fig.~\ref{fig:compoundfig}. 

%\emph{Noise discussion}

We observe several features. First. we observe that there is no measurable difference between the FFTS readout traces and the IQ readout traces. Since the latter is limited by the setup, we conclude that the FFTS readout system does not limit the system noise. 
Secondly, both setups show a large excess noise when probing the MKID 8~kHz off resonance, which is rolled-off at $\sim$15~kHz. This is the magnitude representation of the MKIDs excess frequency noise~\cite{Gao07}, rolled off at the ring response frequency of the resonator (bandwidth $\delta f=f_0/(2Q)\sim 15$~kHz for the MKIDs presented here). Above this roll-off the noise returns to the system noise floor. 
Thirdly, we observe for both datasets a much lower, almost white noise spectrum when probing the MKIDs at 1~MHz off resonance. 
This noise level is consistent with the system noise temperature. The small difference at the highest frequencies between the 1~MHz and 8~kHz off resonance data is identical to the difference in $S_{21}$ between the 8~kHz and 1~MHz off resonance transmission, a direct consequence off our normalization to the carrier amplitude. 

From the measured noise trace at 8~kHz, temperature dependent response of the MKID and in combination with a separate measurement of the quasiparticle lifetime ($\sim$~1~ms) we can estimate a dark NEP~\cite{note} of $\sim 1 \times10^{-17}$W/Hz$^{1/2}$ at 20~Hz.

%\emph{need more in next? ADC figures comes out of nowhere? Is this strong enough on the expandability of technique}
The propose readout of 8 pixels can easily be expanded to much more pixels. With the existing boards the fundamental limit is given by the 8192 bins of the FFTS. However, for so many pixels careful control is required of the peak voltage at each stage in the readout chain~\cite{MKIDcam} to prevent clipping at the ADC and to control intermodulation products from the non-linearity of the amplifiers when overdriven.
For a signal consisting of $n$ tones with identical rms power $P$, the peak power can be approximated by $P_{peak} \sim P n C$, where $n$ is the number of tones and $C$ the Crest factor, where $C=P_{\mathrm{peak}}/nP \sim 25$ using random tone phase. 
The consequence is that the power gain available between the MKIDs and ADC $\propto n$. 
Since the contribution of the ADC to the system noise (referred to the input of the MKIDs) is dependent on the system gain, increasing the amount of probe signals requires a lower noise from the ADC~\cite{MKIDcam} which can be expressed in terms of the bit noise, Effective Number Of Bits (ENOB).
For example, consider a 1000 MKID pixel array with a Q factor of 50000 and probe signal power per MKID of -85~dBm. This would require an ADC with $\sim$~9 ENOB to give the effective noise at the chip level identical to the low noise amplifier noise floor.
In the case of phase or $|S_{21}|^2$ readout, the system noise is higher due to the excess device noise, so relaxing the requirements for the ADC in the example given to about 7~ENOB.

The readout described in this letter gives out $|S_{21}|^2$ response of a MKID. 
A readout in the complex plane is preferred~\cite{Jochem08} giving the phase and amplitude change of the probe signals, enabling either a more linear response (phase readout) or a higher sensitivity for low background applications using high Q MKIDs from both the phase and amplitude of $S_{21}$.
The current FFTS does internally have the complex information from which the complex transmission can in the future be extracted.

In conclusion, we present a MKID readout concept which will facilitate the application of MKIDs to a wide range of applications.
The electronics, as presented here, can already handle $\gtrsim 100$ pixel readouts for ground-based mm/sub-mm astronomy, easily expandable with wider bandwidth tone generation schemes or closer readout packing of the MKIDs. 
This can then be expanded with a phase/amplitude readout to reach the more stringent requirements for very low NEP applications like space born sub-mm spectroscopy, for example SPICA~\cite{SPICA}.

The authors would like to thank Dennis van der Loon, Jan-Rutger Schrader, Rami Barends, Teun Klapwijk and Henk Hoevers for discussions and help.


\begin{thebibliography}{00}
\bibitem{note:bw} The effective noise bandwidth is 1.16$\times$ the binbandwidth~\cite{Klein08}, so spectral leakage is minimized but adjacent bins will have some cross talk particularly if the probe tone falls on the edge of the bin.
\bibitem{note:LED}The MKIDs will respond only very weakly to the optical photon flux since the response is due to direct absorption by the metal film with no optimization for absorption. But the LED illumination allows a sufficient photon flux to be detected.
\bibitem{note} The $|S_{21}|^2$ is a mix of phase and radius response~\cite{Jochem08}, so responsivity is a complicated function, but can be solved exactly from~\cite{Mazin}. For an (under)estimate, use that $\partial |S_{21}|^2/\partial n_qp\sim \partial |S_{21}|^2/\partial f \times \partial f/\partial N_{qp}$, so a dark NEP can be calculated in a similar fashion to~\cite{Jochem08}.
\end{thebibliography}
\end{document}